**Helping students apply superposition principle in problems involving spherical, cylindrical and planar charge distributions**

Jing Li[1], Alexandru Maries[2] and Chandralekha Singh[1]

[1]*Department of Physics and Astronomy, University of Pittsburgh, Pittsburgh, PA 15260*

[2]*Department of Physics, University of Cincinnati, Cincinnati, OH 45221*

**Abstract:** We describe student difficulties in applying the superposition principle in combination with Gauss's law. We addressed these difficulties by developing a tutorial that uses guided inquiry. Students who used this tutorial following lecture-based instruction performed significantly better on these topics than those who did not. Instructors can assign the tutorial as classwork or homework.

**INTRODUCTION**

Gauss's Law is very challenging for students and using it in combination with the superposition principle only adds further layers of complexity. Understanding Gauss's Law requires understanding of the concepts of both electric field and electric flux, and prior work has uncovered some of the common difficulties students have with these concepts. For example, Li and Singh [1] investigated student difficulties with the electric field and the superposition principle and used these difficulties as resources for developing a tutorial to address these difficulties. The difficulties described include confusing electric field and electric force, thinking that only the nearest point charge contributes to the electric field at a specific point, confusion due to the electric field line representation and interpretation of how to use it, and others. Campos et al. [2] investigated the difficulty involved in usefully interpreting the electric field line representation to reason about the electric field due to multiple charges, and inferred from their findings that the use of this kind of representation may actually be detrimental with regard to helping students develop a functional understanding of the concept of electric field. The student difficulty associated with thinking that only the nearest charge contributes to the electric field at a specific point can also be found in a study by Garza and Zavala [3]. They found, for example, that many students thought that a charged vertical bar would "cancel" the electric field created by the point charge on its other side. Student difficulties with the electric field and superposition principle have also been described in the context of continuous charge distributions [4] and were addressed in a tutorial. Student difficulties with electric flux and Gauss's law (and symmetry) have also been investigated [5-8] and used to develop both a 25-item multiple choice test on these concepts [6] as well as a tutorial to address the common student difficulties [7]. Other investigation shows that Gauss's Law is so challenging that students in an upper-level electricity and magnetism course often continue to struggle with using it to reason about the electric field caused by continuous charge distributions [8].

Here, we describe student difficulties and how they were used as a guide for the development and evaluation of a tutorial to help students learn to apply the superposition principle in problem contexts involving spherical, cylindrical and planar charge distributions. The types of problems students were asked to solve using the superposition principle to find the net electric field involved complex situations such as "two adjacent spheres with charge uniformly distributed over their surfaces" or "an infinite cylinder with uniform surface charge adjacent to a point charge". Students struggled with these problems in which Gauss's law is not useful to find the electric field due to the entire charge distribution in "one go".

In Secs. II-IV, we describe student difficulties identified in calculus-based introductory physics and how the tutorial was designed to address them, followed by implementation results (comparing tutorial and comparison groups) and suggestions for how the tutorial could be incorporated in teaching. Details related to development, validation and implementation of the tutorial can be found in Ref. [1] and supplementary materials [9] (which also includes the entire tutorial, pre-test and post-test). Analysis of these difficulties include more than 100 pre-tests and more than 250 post-tests, in addition to individual interviews with a subset of students using a think-aloud protocol about these topics. The interviews were conducted to better understand student difficulties and inform the development of the tutorial, pre-test and post-test.

The analysis of the pre-test and post-test data are part of a systematic study on student difficulties with these concepts and approaches to mitigate them.

## STUDENT DIFFICULTIES AND HOW THE TUTORIAL ADDRESSED THEM

Before we discuss the common difficulties identified, we note that while the difficulties experienced by all students were similar in nature, students in the sections of the class in which the tutorial was used (tutorial sections) were less likely to have difficulties compared to students in the sections that did not use the tutorial (comparison group). In our discussion of these student difficulties, we will refer to specific questions from the pre-test and post-test and specific sections of the tutorial administered to students, all of which are provided in the supplementary material [9].

Many students had great difficulty applying both Gauss's law and the principle of superposition to calculate the vector sum of the electric fields when Gauss's law can be used to find the electric field due to part of the charge distribution. This is despite the fact that on both the pre-test and the post-test, students were given the results from applying Gauss's law to calculate the magnitude of the electric field due to a uniformly charged sphere, line, or plane. Here we give examples of the common student difficulties in correctly applying the superposition principle in the contexts of the problems posed and how the tutorial addressed them.

***Trying to apply Gauss's law for multiple objects instead of using the results from Gauss's law separately for each and using the superposition principle:***

Fig. 1 shows the geometry used in two of the pre-test questions [9]: a thin line of charge parallel to a hollow non-conducting cylinder, both uniformly charged, infinite in extent and with the same charge per unit length. Students are asked to find the direction and magnitude of the net electric field at points A and B. Students can first find the electric field at points A and B due to the line charge and cylinder of charge separately. Then, by using the superposition principle, one can find the net electric field at each of those points. After traditional instruction, on this pre-test question, about 35% of all students found the direction of the electric field at point A correctly while only 14% of them obtained the correct direction of the field at point B. Some students were confused about the difference between the net electric field and its components, correctly finding the component due to each object but unable to combine them to find the net field, and interviews suggested that in some cases it was because they were attempting to apply Gauss's law to the entire charge distribution. In other words, they did not see the relevance of the electric field components to finding the net electric field because they expected that Gauss's law will give them the net electric field for the entire distribution. Similarly, on the pre-test question (3) [9], which described a situation with two uniformly charged non-conducting spheres close to each other, some students drew the field vectors due to each individual sphere, but did not draw the direction of the net electric field.

To address these difficulties, the tutorial presents a situation with two adjacent hollow non-conducting spheres that both carry charges $+Q$. Students are asked to find the magnitude of the electric field at four points, three of which are outside either of the spheres. They are guided to understand that one should use the results from Gauss's law to find the electric field separately due to each sphere at the indicated points and then add the two vectorially. Students are then asked to apply this idea to various situations, helping solidify these concepts [9]. The tutorial also includes dialog between two hypothetical students who discuss the field inside an infinite non-conducting uniformly charged hollow cylinder with a point charge located outside the cylinder. In this scenario, one student insists that the electric field is zero inside the cylinder because no charge is enclosed, while the other states correctly that while the electric field due to the infinite cylinder is zero at that location, the electric field due to the point charge is non-zero, yielding a non-zero net field [9].

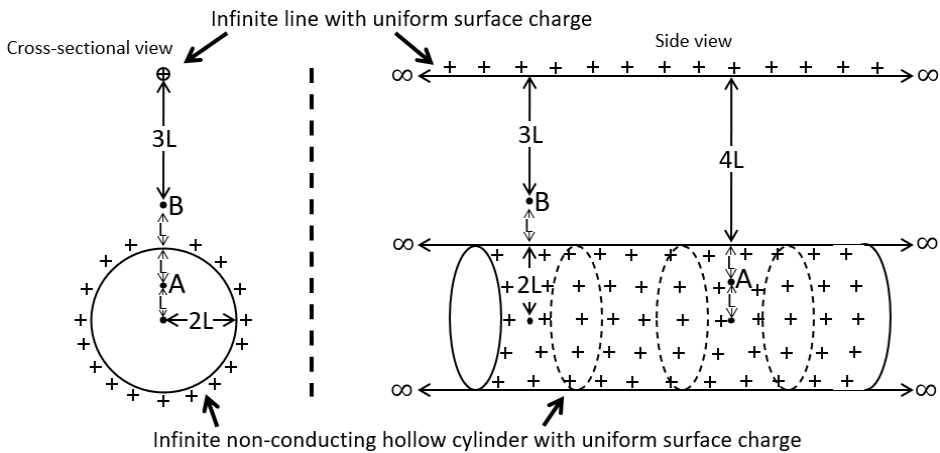

**Figure 1.** Diagram provided for question (1) on the pre-test [9].

*Not considering direction when adding two electric field vectors:* Another difficulty with the superposition principle was assuming that the magnitude of the net field is the sum of the magnitudes of the field due to various objects on which charges were distributed. For example, on pre-test question (2) which uses the situation shown in Fig. 1 [9], some students simply added the magnitudes of the electric field due to the infinite non-conducting cylinder and the infinite line. One interviewed student who used this incorrect method was asked explicitly if the electric field is a vector. The student responded that the electric field is a vector but since he was asked for the magnitude of the net electric field, he added the magnitudes of the electric field contributions coming from the two charged objects. This difficulty might at least in part be due to the fact that using Gauss's law in addition to superposition principle adds to students' cognitive load, and it illustrates the difficulty students have in transferring knowledge of vector addition from mechanics to electrodynamics. It might also be that when using Gauss's law by itself (without superposition), students rarely use the vector nature of the electric field as explicitly as when they do it while first learning to use superposition. Thus, the tutorial helps students recognize that adding electric fields requires using vector addition in situations that require using both Gauss's law and superposition.

To address this difficulty, in the tutorial, students are asked to calculate the magnitude of the net electric field at a point directly in the middle of two uniformly charged spheres as well as at other points where the two electric fields created by the spheres are neither parallel nor opposite [9]. Being asked about the net electric field at a point directly in the middle of the two spheres is intended to elicit student reasoning about direction because at that point, the net electric field is zero (due to the directions being opposite) and students generally recognize this. Additionally, there are other situations in the tutorial in which the vector nature of the electric field is necessary to consider.

*Electric field for multiple parallel infinite sheets with uniform charge is necessarily zero "outside" the sheets:* Post-test questions (4) and (5) [9] describe three infinite parallel sheets with uniform surface charge densities of $+\sigma, -\sigma, +\sigma$. The negative sheet is halfway between the two positive sheets. Here, the most common mistake, especially in the comparison group, was assuming that the electric field is zero at a point outside of the arrangement. For example, one interviewed student made a generalization from the case of a parallel plate capacitor to claim that the field at that point is zero. This type of difficulty could also be related to an inappropriate generalization of what students learn about conductors. When asked explicitly to show why the field is zero using the superposition principle, the student was uncomfortable using this principle and stated that he did not know how to show that the net field is zero but he remembered from class that the net field should be zero outside the parallel plates. The tutorial addresses these types of difficulties with a scenario involving two parallel infinite sheets with uniform charge [9].

Apart from these difficulties with the superposition principle, below we categorize some common conceptual issues with finding the net electric field.

***Measuring distances from the surface of a uniformly charged sphere or cylinder to find the field:*** At the beginning of the pre-test and post-test, students were explicitly given the magnitude of the electric field due to a sphere of charge at a distance $r$ from its center. However, some students appear not to have taken note of this instruction carefully. One common difficulty was taking ***r*** as the distance measured from the surface. Students had similar difficulties with charged cylinders. During interviews, even though students were told that the infinite line and the infinite cylinder in Fig. 1 have the same charge per unit length, many still claimed that the cylinder would have more effect at point B because one side of the cylinder is closer to the point where the electric field is being calculated.

This difficulty was addressed in multiple contexts in the tutorial by asking students questions in which knowing that the distances should be measured from the center of the sphere/cylinder was necessary. For example, the tutorial had a question very similar to question (1) on the pre-test (Fig. 1). Students had to find the net electric field due to a point charge and an infinitely long cylinder where they needed to measure the distance from the center of the cylinder [9]. The tutorial also included a discussion between two hypothetical students in which one voices the common difficulty (measure the distance from the surface) and the other voices the correct approach [9]. Also, in the tutorial scenario with two spheres described earlier, students have to evaluate the validity of hypothetical student conversations focusing on these issues [9].

***Claiming that a non-conductor shields the inside from the electric field due to external charges:*** For example, on pre-test question (1) [9], some students claimed that the infinite line with uniform charge in Fig. 1 cannot produce a field at point A inside the infinite non-conducting hollow cylinder because point A is inside the cylinder and is shielded from charges outside. Another example comes from question (3) on the pre-test [9] which described a situation in which two hollow non-conducting spheres with uniform charge distributions on the surface are near each other. When asked to determine the net electric field at points inside of one of the spheres (not at the middle), some students claimed that the field is zero everywhere inside the sphere and argued that the second sphere cannot produce a field inside the first sphere because of shielding. This notion of shielding was retained by the students despite being reminded during the interviews that the object on which the charges are distributed is non-conducting and there are also charges outside in the problem posed. For example, on post-test question (1) [9], in which a hollow, non-conducting, uniformly charged sphere is near a point charge and asks students to draw vectors showing the electric field at a point inside the sphere, some students said the electric field inside the sphere is zero. When during interviews, they were explicitly asked by the interviewer why the point charge near the charged sphere does not produce a field inside the hollow sphere, some students referred to the shielding of the inside of the sphere from the charges on the sphere and the charges outside of the sphere. Some students noted that they could not explain exactly why the non-conducting sphere will produce shielding, but that they remembered that the electric field must somehow cancel out in the hollow region for all shapes and charge distributions. Further prodding suggested that due to a lack of thorough understanding, these students were often overgeneralizing or confusing two different facts: the symmetry argument that shows (using Gauss's law) that the electric field for a sphere with a uniform surface charge is zero everywhere inside regardless of whether the sphere is conducting or non-conducting, and/or the fact that the electric field inside a conductor in equilibrium is always zero regardless of the shape of the conductor.

This difficulty is also addressed in multiple contexts in the tutorial. For example, students are asked to calculate the electric field in a situation similar to the one shown in Fig. 1 (at point A) [9]. Another example of addressing this difficulty is asking students to contemplate a discussion between two students in which one insists that the electric field is zero inside of the cylinder (for a situation similar to the one described in Fig. 1 at point A) while the other correctly states that the charge distribution on the cylinder does not create an electric field inside of the cylinder, but the outside point charge does [9]. Additionally, students have to use this concept both for this situation and another similar one (two hollow non-conducting spheres near each other) to find the magnitude of the field inside one of them [9].

**PERFORMANCE OF THE TUTORIAL AND COMPARISON GROUPS**

The tutorial along with the pre-test and post-test are provided in the supplementary material [9]. The pre-test and post-test results show that the tutorial was effective in helping students apply the superposition principle correctly in these contexts. They are described in detail in the supplementary material [9]. Students in the comparison group showed a slight improvement from 24% on the pre-test to 35% on the post-test on average, whereas for the tutorial group, students improved from an average of 27% to 81%.

**IMPLICATIONS FOR TEACHING**

We used the research on student difficulties as a guide to develop and evaluate a tutorial to help students learn to apply Gauss's law along with the superposition principle correctly. Results suggest that the tutorial is effective at helping students apply the superposition principle in these contexts.

The tutorial can be used in many ways in its entirety or in part as follows:

- The tutorial could be used in part or in full either in class or in recitation. We note that the tutorial is designed to be used by the instructors as a supplement, not as the first tool to introduce students to these concepts. When students work in small groups on the tutorial, the instructor or the teaching assistant can move around to ensure that students engage productively with each other and make sense of the guided inquiry-based sequences. Also, parts of the tutorial could be used if there isn't enough time for students to work all the problems. For example, the tutorial discusses three different situations with uniform surface charge distribution: two spheres, a cylinder and point charge, two large sheets. One can select any of these three scenarios and have students work on it before having a related class discussion.

- After students learn about Gauss's law, questions in the tutorial could be framed into think-pair-share or clicker questions. These two approaches are similar except think-pair-share does not require feedback via clickers and questions using think-pair-share may or may not be framed in a multiple-choice format. However, even in the think-pair-share format, students can be asked to share what they discussed with their peers followed by a full class discussion. For example, the tutorial involves a situation in which there is an infinitely long uniformly charged non-conducting hollow cylinder close to a point charge. Students could be presented with just the non-conducting cylinder first and asked about the electric field inside the cylinder (zero) possibly with several options to select from. Then, the point charge could be added to the situation and students could be asked a similar question again for both objects together. Similarly, students could be asked about the field outside a cylinder or a sphere in order to help them measure distance correctly before adding a second charged object. This approach helps gradually add complexity and could help manage students' cognitive load. The support can be gradually reduced for other tutorial questions and students can be asked directly about the electric field due to objects together to help them develop self-reliance.

- The tutorial includes multiple conversations between hypothetical students in which one student voices a common difficulty and the other provides a correct explanation or other conversations in which they both provide parts of a correct explanation. These conversations can also be used in class as think-pair-share or clicker questions asking students about which hypothetical student they would agree with, e.g., student 1, student 2, neither, or both. This can be followed by class discussion about the underlying concepts.

- The tutorial can alternatively be assigned as homework after students have had prior instruction in relevant concepts. When assigned as homework, after students submit it as homework, the tutorial problems can be discussed briefly in class.

Thus, the tutorial is designed to allow instructors flexibility in using their preferred methods for initial instruction, ranging from traditional lectures to assigned reading followed by student questions and in-class discussion. If only part of the tutorial is used by instructors, they could select the pre-test and post-test questions that correspond to the material in those sections.

In this manner, instructors can use either the full or part of the tutorial based upon their preference as a resource inside or outside of the classroom to actively engage students in learning these challenging concepts.

## ACKNOWLEDGEMENTS

The authors would like to thank Z. Isvan for help in tabulating data and F. Reif, P. Reilly, and R. P. Devaty for helpful discussions. The authors also thank the faculty and students who helped with this study.

# SUPPLEMENTARY MATERIAL

# METHODOLOGY USED IN THE STUDY

The students who participated in this study were enrolled in different sections of a second semester calculus-based college introductory physics course at a large research university, mainly taken by engineering (70% of the students are engineering majors), chemistry, mathematics and physics majors. This course covers E&M and wave optics. It is taken after the first introductory physics course, which covers mechanics and waves. The majority of the several hundred students in the different sections of this course were first year students. The instruction for students in this course consisted of four hours of lecture time and one hour of recitation time. The different sections of the course were typically taught by different instructors and the recitation classes were taught by graduate teaching assistants. All of the sections of the course primarily had lecture-based instruction in the four hours of lecture time (except in the sections of the course that we designate the experimental group, in which students worked on the tutorial in one class after lecture-based instruction). In the recitation classes, the graduate teaching assistants answered questions about homework from students and solved example problems on the board on relevant topics.

The reason there are several sections of this same course offered simultaneously at the university where the study was carried out is that this is a required course for all engineering majors and also for other majors, e.g., from chemistry, mathematics, etc. The content covered by all the different sections of the course (both experimental and comparison groups) is the same, with the exception that the tutorial group worked on the tutorial in small groups during one class period, whereas for the comparison group, the instructor solved example problems based upon the tutorial for students. Also, students who did not complete the tutorial during the regular class in the experimental group, were asked to complete the rest of it at home. Each week students were asked to complete homework from the course textbook some of which, e.g., involved finding the electric field at various points due to several thin parallel infinite sheets with uniform surface charge. In all sections of the course, each week, after students submitted the homework on a particular topic, they were given a quiz on that material in the last 15-20 minutes of the recitation class. We note that this investigation employs a quasi-experimental design in that we did not have control over whether a particular student will be in the section of the course in which the tutorial was used and who their instructor and teaching assistant (for the recitations) would be. Also, although students in all sections of the course used the same textbook, we did not have control over the textbook homework assignments or the midterm and final exam given by the instructors of different sections. However, the conceptual survey of electricity and magnetism given to some of the sections in the previous years as a pre-test and post-test suggests that student performance on average in various sections of the course is comparable at the beginning of the course (pre-test) and on the post-test after traditional lecture-based instruction. We note that the development and validation of the guided inquiry-based tutorial was carried out using a method similar to that described in Ref. [4]. Table 1 summarizes this process for data collection before, during and after the development of the questionnaire and different versions of the tutorial similar to Ref. [4].

Table 1. *Data collection protocol before, during and after the development and validation of the questionnaire and the tutorial.*

| Before the development of the questionnaire and tutorial | During and after the development of the preliminary version of tutorial: individual interviews with students and individual discussions with faculty members |
|---|---|
| • Cognitive task analysis from an expert perspective of the relevant concepts<br>• Development of questions and discussions with faculty teaching this course about them to validate them<br>• Several years of examining student responses to<br>  ○ open-ended questions<br>  ○ multiple choice questions<br>• Individual interviews with students (N=5) | • One-on-one interviews with students (N=10) examining responses to<br>  ○ tutorial pre-test questions<br>  ○ inquiry-based learning sequences in tutorial<br>  ○ tutorial post-test questions<br>• Discussions with faculty members<br>• Refinement based upon feedback from students and faculty |

Table 2. *Summary of activities in the experimental group and comparison group in different sections of the course. $N_i$ is the number of students in each group.*

| | Learning activities before the pre-test | Pre-test (quiz graded for completeness) | Learning activities after lecture-based instruction and pre-test | | Post-test (quiz graded for correctness) |
|---|---|---|---|---|---|
| **Tutorial group or experimental group** | Traditional lecture-based instruction in relevant concepts for both groups. | - $N_2 = 57$ - | Students in the tutorial group worked on the tutorial during a class in small groups. | Recitation classes were similar for both groups and students had an opportunity to ask questions about textbook homework problems (some of the textbook problems, e.g., involved finding electric field due to several parallel infinite uniform sheets of charge). | $N_1 = 87$ $N_2 = 57$ $N_3 = 65$ |
| **Non-tutorial group or Comparison group** | | $N_4 = 57$ | Instructor worked out problems similar to the ones in the tutorial on the board in class. | | $N_4 = 57$ |

Table 2 is a summary of activities in the experimental group and comparison group (see left to right for chronology) in various sections of the course in the semester in which the final version of the tutorial was administered in-class in some sections of the course. For all sections in Table 2, $N_1$, $N_2$, $N_3$, and $N_4$ refer to the number of students in each section for which the pre-test or at least the post-test data were available. As can be seen in Table 2, both groups had traditional lecture-based instruction in the relevant concepts. After traditional lecture-based instruction, students took the pre-test in class as a quiz which was graded for completeness. Then, during class, students in the experimental group worked on the tutorial in small groups. In the comparison group, the instructor worked out problems similar to the ones in the tutorial while discussing concepts. Students in both groups were required to complete the same textbook homework. Recitation classes were similar for both groups and students had an opportunity to ask questions about textbook homework problems (some of the textbook problems, e.g., involved finding electric field due to several parallel infinite uniform sheet of charge). Then, the post-test was administered as a quiz at the end of the recitation. As shown in Table 2, in the tutorial group, the number of students who were administered the pre-/post-test ranged from 57-87 (two of these sections did not take the pre-test due to time-constraints). The post-test was returned after grading (unlike the pretest).

## TUTORIAL: Revisiting the Superposition Principle after Gauss' Law

- Assume that all non-conductors (insulators) are non-polarizable.

In the following section, we will use the principle of superposition to find the electric field at a point due to multiple sources. Remember, the electric field at a point due to different sources of charges can be added vectorially to find the net field.

(I) Two identical hollow non-conducting spheres, each of radius R and the same total charge +Q uniformly distributed on the surface, are a distance 4R apart measured center to center, as shown below. You are asked to find the electric field at points A, B, C, and D which are in the plane shown in the figure. Points A, B and C are on the straight line joining the centers of the two spheres and:
- A is to the left of sphere 1 at a distance 2R from its center.
- B is equidistant from the two spheres on the straight line joining their centers.
- C is at the center of sphere 2.
- D is on the perpendicular bisector of the straight line joining the centers of the spheres at a distance 2R above point B.

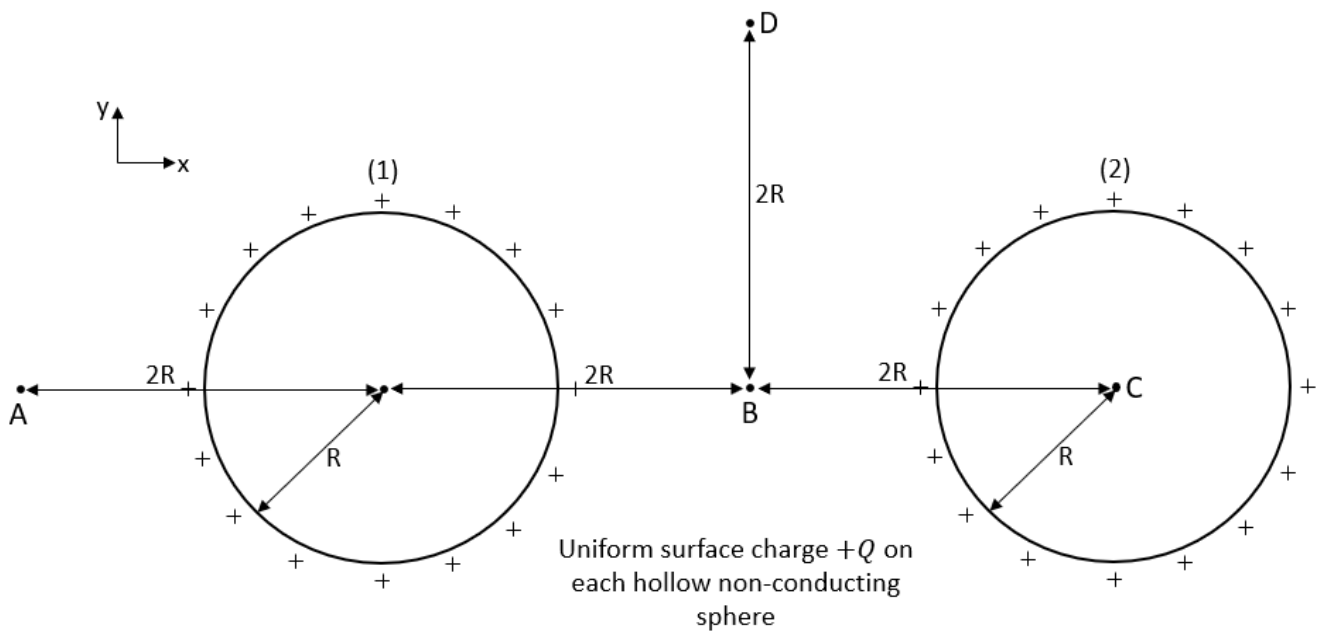

Figure 1. A schematic diagram for situation (I)

(1) Consider the following plan proposed by John:
- Step 1: First, think about each sphere one at a time, in isolation. Since the charge is uniformly distributed on the surface, use Gauss's Law to determine the electric field at each point due to the charges on each sphere.
- Step 2: Use your results from Step 1 and the principle of superposition to calculate the net field at each of the points A, B, C and D.

Is this a reasonable plan? Explain.

(2) Use the principle of superposition (the net electric field is the vector sum of the electric field due to various sources) proposed by John to calculate the magnitude and direction of the net electric field at points A, B, C and D assuming that the magnitude of the electric field due to an isolated sphere of radius $R$ with charge $+q$ uniformly distributed on its surface is:
- $|\vec{E}| = 0$ at any point inside
- $|\vec{E}| = kq/r^2$ at any point outside a distance $r$ away from the center

(3) Consider the following conversation between John and Susan:
- John: "For points outside a sphere, the distance $r$ in $|\vec{E}| = kq/r^2$ should be measured from the surface of that sphere."
- Susan: "I disagree. For a point outside the sphere, the distance $r$ should be measured from the center of that sphere because the uniform sphere of charge behaves as a point charge at the center."

With whom, if either, do you agree? Explain.

(4) Is the net electric field zero at any of the points? Explain.

(5) Which points, if any, have net electric field that is along the $x$ axis? Which points, if any, have electric fields that are along the $y$ axis? Make sure to draw arrows to show the directions of the net electric field at each of the points shown in the figure above.

(II) Consider the following situation in which there is a point charge $+Q$ near an infinite non-conducting cylinder with uniform surface charge (charge per unit length $\lambda$). Points A, B and the point charge $+Q$, are all in the same cross-sectional plane of the cylinder and all are on a straight line joining the point charge $+Q$ to a point on the cylindrical axis.

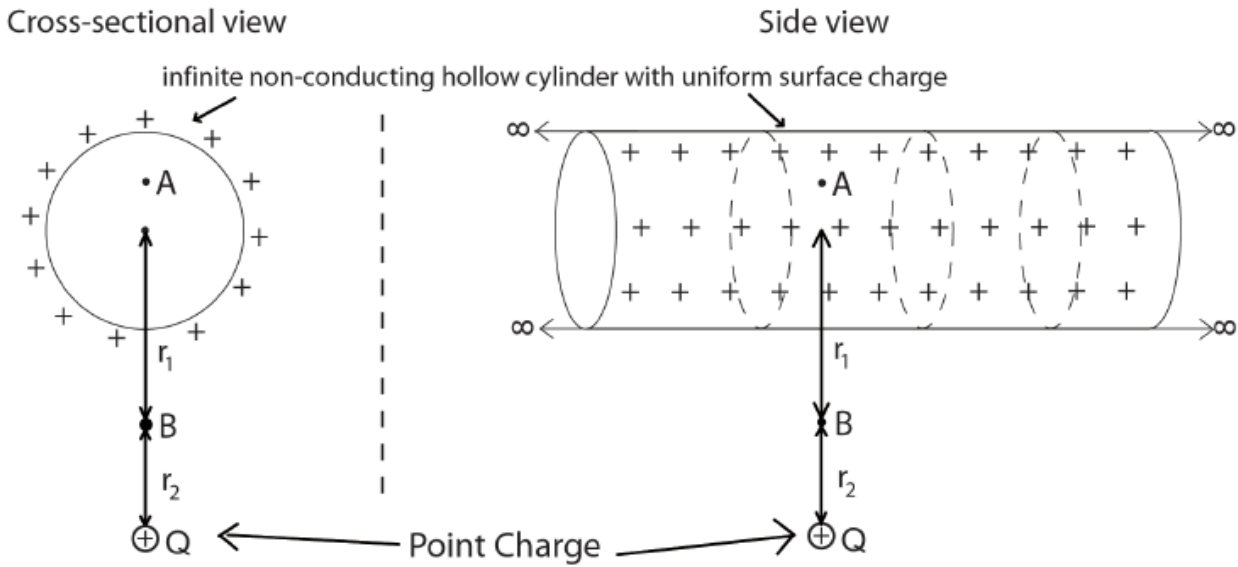

Figure 2. A schematic diagram for situation (II)

(1) Consider the following conversation between Harry and Susan:
- Harry: "The net electric field should be zero everywhere inside the hollow region of the cylinder since there is no charge enclosed".
- Susan: "I disagree. The electric field due to the infinite cylinder of charge is zero inside it but not the electric field due to the point charge $+Q$. Therefore, the net electric field inside the cylinder cannot be zero!"

Explain why you agree or disagree with each.

(2) Consider the following conversation between Pria and Mira:
- Pria: "The point charge and the infinite cylinder of charge together do not have high enough symmetry so that it is not useful to construct one Gaussian surface and exploit Gauss's law to find the magnitude of the net electric field, $|\vec{E}|$, at a point due to both."
- Mira: "I agree. But we can use the principle of superposition here to find the electric field at a point. We know the field at each point due to a point charge using Coulomb's law and due to an infinite cylinder of charge using Gauss' law. All we need to do is to add them vectorially."

Explain why you agree or disagree with each.

(3) In the figure above, point A is inside the infinite non-conducting cylinder of charge. Draw separate arrows to indicate the directions of the electric field at point A due to the point charge $+Q$ and due to the infinite cylinder of charge. If the field is zero due to either of these, say so explicitly.

(4) Is there a net electric field at point A? Make sure your answer is consistent with your answer to question (1). If there is a net electric field at point A, draw an arrow in the cross-sectional view above to show its direction.

(5) In the figure above, point B is outside the infinite cylinder of charge. Draw separate arrows to indicate the directions of the electric field at point B due to the point charge $+Q$ and due to the infinite cylinder of charge. If the field is zero due to either of these, say so explicitly.

(6) Consider the following conversation between Pria and Mira:
- Pria: For point B outside the cylinder, the distance $r$ in $|\vec{E}| = \lambda/(2\pi\varepsilon_0 r)$ due to the cylinder should be measured from the surface of the cylinder.
- Mira: I disagree. For a point outside the cylinder, the distance $r$ should be measured from the axis of the cylinder because the uniform cylinder of charge behaves as a line of charge at the axis.

With whom, if either, do you agree? Explain.

(7) If the radial distance of point B from the center of the cylinder is $r_1$ and from the point charge $+Q$ is $r_2$, what is the magnitude of the electric field at point B due to the point charge and due to the cylinder of charge separately?

(8) Suppose the magnitude of the electric field at point B due to the point charge $+Q$ is greater than that due to the cylinder of charge. Using this assumption, draw an arrow to show the direction for the net electric field at point B above.

(9) Suppose the magnitude of the electric field at point B due to the point charge $+Q$ is exactly the same as that due to the cylinder of charge. What will be the direction of the net electric field at point B then?

(III) Two parallel thin infinite sheets with uniform charge are a distance $d$ apart. The charge density (charge per unit area) on sheet (1) is $+\sigma$ and on sheet (2) is $-\sigma$. Find the magnitudes of the net electric field at points A, B, and C in regions (I), (II) and (III) shown below due to the charges on both infinite sheets.
(Note: The magnitude of the net electric field due to a single infinite uniformly charged sheet is $\sigma/(2\varepsilon_0)$ and the direction of the field is perpendicular to the sheet everywhere.)

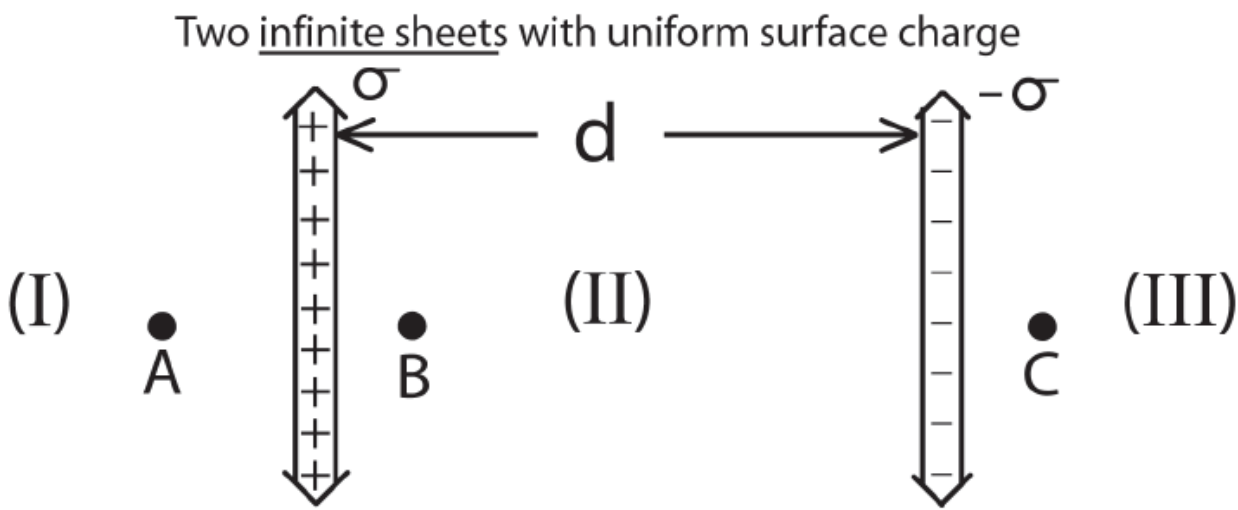

Figure 3. A schematic diagram for situation (III)

(1) Draw arrows labeled $\vec{E}_1$ above to show the direction of the electric field due ONLY to the infinite sheet with positive charge at each of the three points. (Hint: The electric field due to the positively charged sheet should point away from the sheet everywhere.)

(2) Draw arrows labeled $\vec{E}_2$ above to show the direction of the electric field due ONLY to the infinite sheet with negative charge at each of the three points. (Hint: The electric field due to the negatively charged sheet should point towards the sheet everywhere.)

(3) Consider the following conversation between Pria and Mira:
- Pria: "In region I, the length of the arrow for $\vec{E}_1$ should be longer than the length of the arrow for $\vec{E}_2$ because the electric field due to the positively charged sheet will be stronger than the field due to the negatively charged sheet."
- Mira: "I disagree. Since the sheets are infinite, the magnitude of the electric field everywhere due to each sheet is the same regardless of how far you are from its surface."

With whom, if either, do you agree? Explain and draw the lengths of the arrows above to reflect your answer.

(4) Use the superposition principle to find the magnitude of the net electric field at each of the three points.

(5) Consider the following conversation between Pria and Mira:
- Pria: "The net electric field at points A and C is zero because the field due to each sheet has equal magnitude but opposite direction. The net electric field at point B is $\sigma/\varepsilon_0$ and it points away from the positive sheet."
- Mira: "Yes. In fact, the magnitude of the net electric field at any point in regions (I) and (III) will be zero and in region (II) will be $\sigma/\varepsilon_0$"

With whom, if either, do you agree? Explain.

**PRE-/POST-TESTS**

The following information was provided for both the pre-/post-tests.
- For all problems below, leave your answers in terms of $\varepsilon_0$.
- Assume all insulators (non-conductors) are non-polarizable.
- The magnitude of the electric field due to an infinite uniform line of charge with charge density $+\lambda$ at a distance $r$ from the line is $+\lambda/(2\pi\epsilon_0 r)$.
- The magnitude of the electric field due to a uniform sphere of charge, with total charge $+Q$, outside the sphere at a distance $r$ from the center is $k\,Q/r^2$ where $k = 1/(4\pi\epsilon_0)$
- The magnitude of the electric field due to a single uniformly charged infinite sheet with surface charge density $+\sigma$ is $\sigma/(2\varepsilon_0)$.

**PRE:** (1) Consider a thin line and a hollow non-conducting cylinder of radius $2L$, both infinite in extent with charge uniformly distributed on them (see the cross-sectional and side view in Fig. 4). Both have the same charge per unit length $+\lambda$. They are parallel to each other. The perpendicular distance (shortest distance) separating the infinite line of charge from the axis of the hollow charged cylinder is $6L$. Points A and B lie on radial straight lines between the line of charge and the axis of the charged cylinder as shown in Fig. 4. On the cross-sectional view in Fig. 4, draw arrows to show the directions of the net electric field at the two points A (inside the cylinder at a radial distance $L$ from the axis) and B (outside the cylinder at a radial distance $3L$ from the axis).

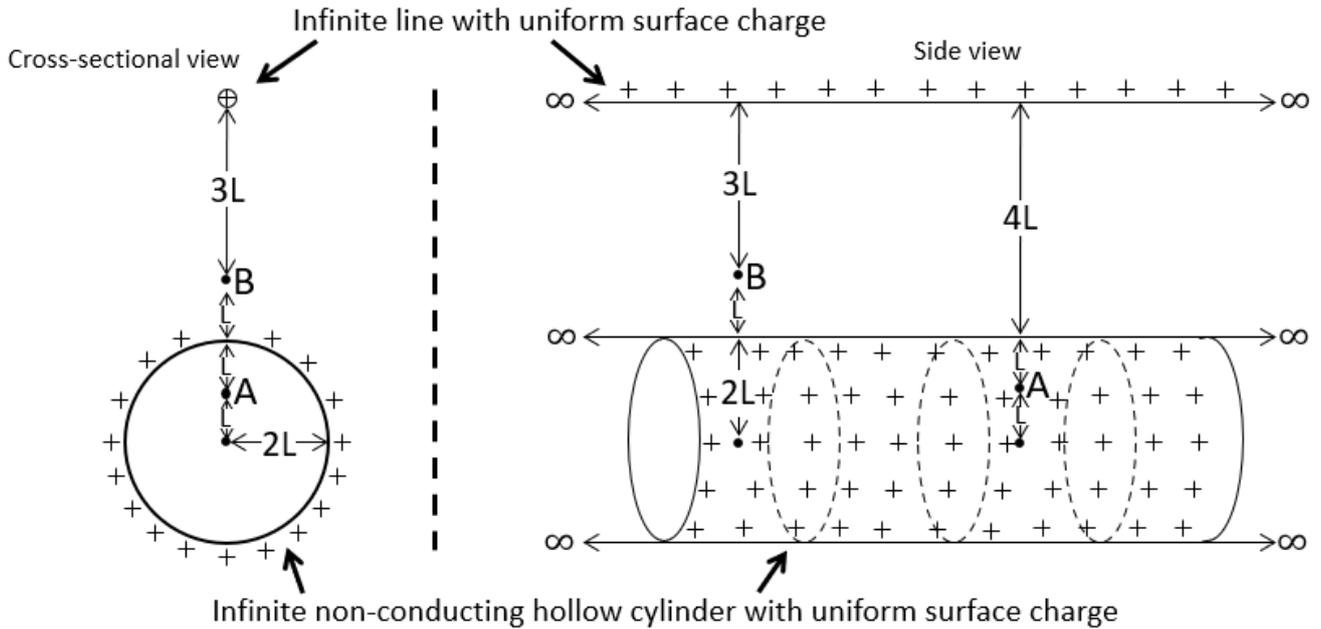

**Figure 4**. A schematic diagram for the pre-test questions (1) and (2)

(2) Write the *magnitude* of the net electric field at points A ($|\vec{E}_A|$) and B ($|\vec{E}_B|$) in Fig. 4 in terms of $\lambda$ and the distances shown.

(3) A cross section through the equators of two adjacent identical hollow non-conducting spheres, each with uniform surface charge $+Q$, is shown in Fig. 5. Find the direction of the net electric field at points A, B and C. Point A is in the hollow region of one of the spheres. Points B and C are on the perpendicular bisector of the straight line joining the centers of the two spheres. Explain your reasoning. Hint: Use Gauss' law and the principle of superposition.

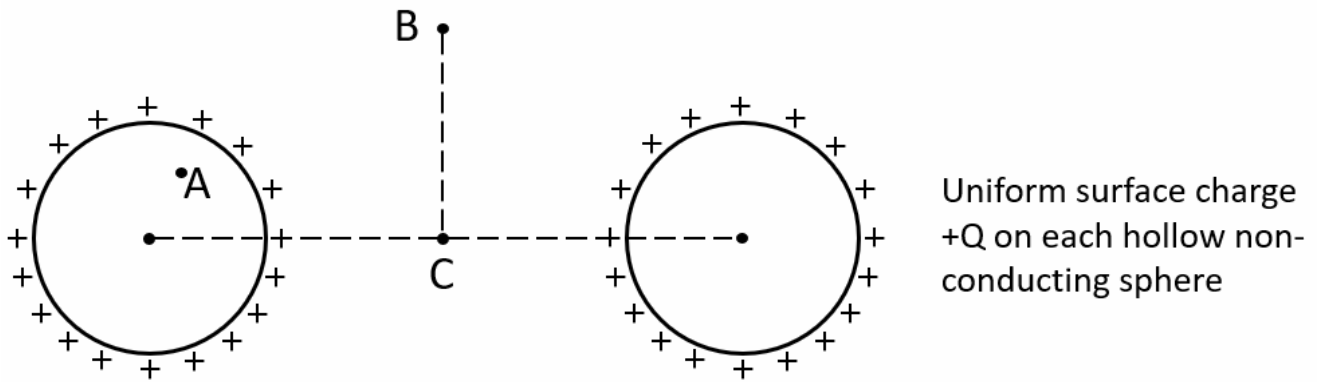

**Figure 5**. A schematic diagram for the pre-test question (3)

**POST**: (1) In Fig. 6 (cross-sectional view), a point charge $+Q$ is near a thin hollow non-conducting sphere of radius $4L$ that has an <u>equal</u> amount of charge $+Q$ uniformly distributed on its surface. No other charges are around. Draw arrows to show the direction of the net electric field at two points: A (inside the sphere) and B (outside the sphere) shown in Fig. 6.

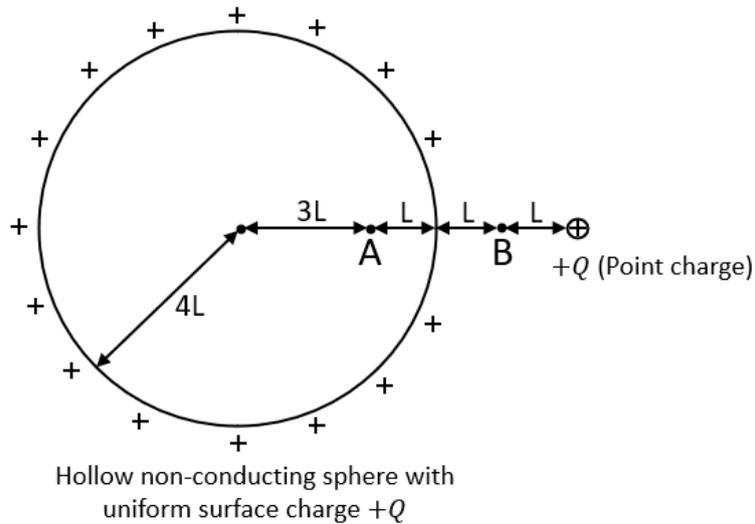

Hollow non-conducting sphere with uniform surface charge $+Q$

**Figure 6.** A schematic diagram for the post-test questions (1) and (2)

(2) Write the *magnitude* of the net electric field at points A ($|\vec{E}_A|$) and B ($|\vec{E}_B|$) in Fig. 6 in terms of charge $+Q$ and the distances shown.

(3) Two adjacent identical hollow non-conducting <u>cylinders</u>, both infinite in extent and each with the same uniform negative charge per unit length $-\lambda$ are shown in Fig. 7. The axes of the two cylinders are parallel. A cross section of the cylinders is shown in Fig. 7. Find the direction of the net electric field at points A, B and C. Point A is in the hollow region of one of the cylinders. Point B is on the perpendicular bisector of the straight line joining the centers of the two cylinders in the plane of the paper.

Explain your reasoning for each point. (Note: If you cannot find the exact direction at any of the three points shown in Fig. 7, draw an approximate direction.)

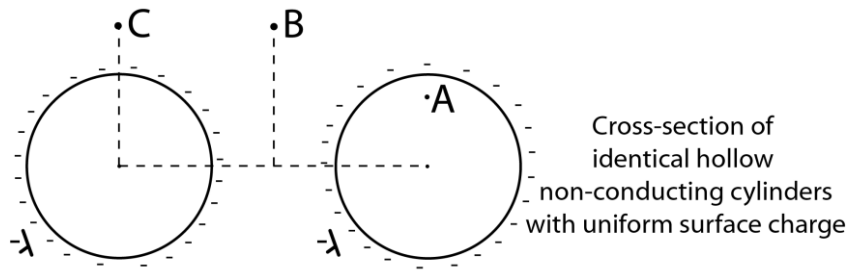

**Figure 7**. A schematic diagram for the post-test question (3)

(4) Three parallel thin infinite sheets with uniform charge are such that the adjacent ones are a distance $d$ apart (see Fig. 8). The surface charge density (charge per unit area) on sheets (1) and (3) is $+\sigma$ and on sheet (2) is $-\sigma$. Find the magnitudes of the net electric field at points A and B shown in Fig. 8 due to the charges on the sheets.

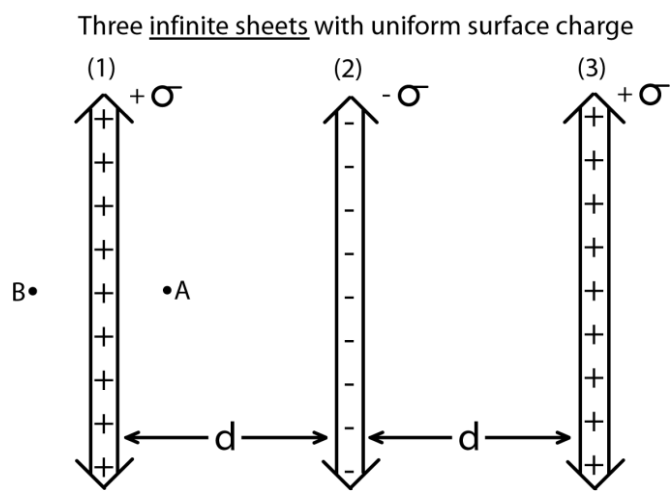

**Figure 8**. A schematic diagram for the post-test questions (4)-(6)

(5) Show with arrows the direction of the net electric field at points A and B due to the infinite sheets of charge in Fig. 8.
(6) How would the electric field at points A and B obtained in questions (4) and (5) be affected if sheets (2) and (3) in Fig. 8 are removed? Explain.

## PERFORMANCE OF THE TUTORIAL AND COMPARISON GROUPS

The pre-test was given immediately after relevant instruction for both groups. Then students worked on the tutorial in class in the tutorial group and instructor worked through problems similar to the tutorial problems in class in the comparison group. The post-test was administered the following week for both groups as part of the weekly recitation quiz after students had the opportunity to complete all the homework problems related to these topics and had opportunity to ask questions about them. The pre-test and post-test were graded by two researchers based upon an agreed upon rubric with inter-rater reliability better than 85%. Based upon whether the student responses were conceptually correct or not, the grading rubric scores each answer as correct or incorrect and if there was an explanation required, student responses for that part of the question were graded on the following scale: full point for correct explanation, no point for incorrect or no explanation and half point for partially correct.

Table 3 shows the average pre-/post-test scores on each question and also overall for three sections of the course in which the tutorial was administered. As shown in Table 3, the pre-test (after lecture-based instruction) was not administered to sections 1 and 3 due to instructor's concerns about time constraints and the instructor only administered the first three questions on the post-test in section 2. Table 3 shows that the average performance was significantly better on the post-test compared to the pre-test for the tutorial group. The differences in the performance of different sections on the post-test shown in Table 3 may partly be due to the differences in student samples, instructor/TA differences or the manner in which the tutorial was administered and are measures of the types of differences instructors (who may use the tutorial in the future) may observe in their classes depending upon the details of the tutorial implementation and their students' prior knowledge.

Table 4 shows the pre-/post-test data for the comparison group which consists of an equivalent introductory physics section of the course in which students did not work on the tutorial. Table 4 shows that the post-test performance of students in the comparison group is only slightly larger than their performance on the pre-test (35% compared to 24%).

The results of a *t*-test that compares the average performance of the tutorial and non-tutorial sections on the pre-/post-tests in Tables 3 and 4 show that the performance of both the comparison and the tutorial group on the pre-test was poor after traditional lecture-based instruction (the two groups show comparable performance with a $p$ value of 0.6). However, the post-test students in the tutorial group significantly outperformed students in the comparison group ($p$ value < 0.001).

**TABLE 3:** *Average total scores and scores on individual questions on the pre-/post-test for the tutorial group. The pre-test was administered after traditional lecture-based instruction but before the tutorial. The pre-test was not administered to students in sections 1 and 3 due to time constraints and only three questions were administered on the post-test to section 2. N refers to the matched number of students in a given section who took both the pre-/post-tests and "total" refers to the total average percentage score including all questions on the pre-test or post-test administered to a given section. The relative weights for pre-test 1, 2, and 3, were 30%, 40% and 30% and the relative weights for post-test questions 1, 2, 3, 4, 5, and 6, were 10%, 20%, 20%, 20%, 10%, and 20%, respectively.*

| Section | N | Pre-test | | | | Post-test | | | | | | |
|---|---|---|---|---|---|---|---|---|---|---|---|---|
| | | 1 | 2 | 3 | Pre-total | 1 | 2 | 3 | 4 | 5 | 6 | Post-total |
| 1 | 87 | - | - | - | - | 69% | 59% | 68% | 69% | 96% | 73% | 70% |
| 2 | 57 | 20% | 26% | 35% | 27% | 82% | 76% | 85% | - | - | - | 81% |
| 3 | 65 | - | - | - | - | 93% | 81% | 90% | 91% | 98% | 92% | 90% |

**TABLE 4**: *Average percentage total scores and scores on individual questions on the pre-test and post-test for a section of the comparison group. The relative weights for the pre-test and post-test questions for calculating the total is similar to that for the tutorial group.*

| N | Pre-test | | | | Post-test | | | | | | |
|---|---|---|---|---|---|---|---|---|---|---|---|
| | 1 | 2 | 3 | Pre-total | 1 | 2 | 3 | 4 | 5 | 6 | Post-total |
| 57 | 18% | 11% | 49% | 24% | 18% | 19% | 44% | 27% | 69% | 34% | 35% |

As a comparison between the tutorial and comparison groups on a specific question, on post-test question (1), 87% of the students in the tutorial group provided the direction of the electric field at points A and B correctly. In the comparison group, however, 46% of the students provided the direction of the field at point A correctly and only 30% of them provided the direction of the field at point B correctly. On post-test question (3), students were given two adjacent identical hollow non-conducting cylinders which were both infinite in extent and each had the same uniform negative charge per unit length. They were asked to draw the direction of the net field at points A, B and C. Students could determine the direction of the field due to each uniformly charged cylinder using Gauss's law and then use vector addition to determine the direction of the net electric field. In the tutorial sections, at least 90% of the students found the directions of the electric field at points B and C correctly but in the comparison group, only 37%-53% of students found the directions of the field at points C and B correctly. In the comparison group, the majority of the students struggled with the correct direction at point A, but even in the tutorial group, only 60% of students found the direction at point A correctly (suggesting that this question is extremely difficult).

On post-test questions (4) and (5) about the electric field at various points due to three parallel infinite sheets with uniform charge, students in the tutorial group performed significantly better (see Tables 3 and 4) than students in the comparison group even though all students were provided the equation for the magnitude of the electric field at a point due to an infinite uniform sheet with surface charge $+\sigma$ ($|\vec{E}| = \frac{\sigma}{2\epsilon_0}$, showing that the field due to an infinite uniform sheet of charge is independent of the distance from the sheet). Many students struggled with the fact that the electric field at a point is independent of the distance of the point where the field is calculated from the infinite uniform sheet of charge in addition to the issues related to applying the superposition principle correctly.